# Fluorescent emission and nondestructive 3D textural imaging in geologic materials by multiphoton microscopy

SAMUEL D. CROSSLEY,[1,*] COLBY L. DONNER,[2] JOSH MAGNUS,[2] LAM NGUYEN,[2] AND KHANH KIEU[2]

[1] *Department of Planetary Sciences and Lunar and Planetary Laboratory, The University of Arizona, 1629 E University Blvd, Tucson, AZ, 85721, USA*
[2] *Wyant College of Optical Sciences, The University of Arizona, 1630 E University Blvd, Tucson, AZ 85721, USA*

**Abstract:** We greatly expand the application of multiphoton microscopy to geological investigations by using a tightly focused femtosecond laser beam to excite fluorescent emissions among minimally prepared rock and mineral samples. This new finding provides a tool for spatially resolving UV-visible fluorescent sources in minerals. Using a unique combination of harmonic generation and fluorescence, we explore applications to mineralogical investigations of terrestrial rocks and astromaterials. We report first-order demonstrations for 3D imaging of fluid inclusions in minerals and radiation-induced luminescence in meteorites. Nonlinear optical mineralogy, enabled by multiphoton microscopy, provides unique insights in mineralogic samples and holds the potential to revolutionize the analysis of geologic and astromaterials samples in the coming years.

## 1. Introduction

Luminescence emission from gems and minerals was first investigated by Stokes in the mid-19th century [1], who established the "Stokes Law of Fluorescence" or Stokes shift when studying the luminescence emission of fluorite when stimulated with ultra-violet light [2]. Fluorescent spectroscopy of minerals has since been utilized to provide useful information on the origin of the luminescence centers arising from chemical/mineral impurities or intrinsic defects in the crystal structures of gems and minerals [2]. Among many fields of study, information about the chemical and structural state of mineral species is critical to investigations in geosciences, economically valuable ore deposit resources, and planetary sciences. The main techniques used to measure luminescence emission from minerals are based on thermoluminescence, photoluminescence, cathodoluminescence, or X-ray sources [2,3]. However, these techniques do not usually provide both quantitative measurements and high spatial resolution for the distribution of luminescent centers within a mineral. Additionally, some techniques that do allow for spatially resolved measurements, e.g. cathodoluminescent or X-ray spectroscopy, only provide emission from the surface of the sample. Multiphoton microscopy (MPM) with a tightly focused femtosecond laser beam is a powerful imaging technique that has been widely adopted in the biomedical sciences [4-7]. In contrast to aforementioned techniques, it enables sub-micron resolution mapping of biological tissues in 3D using second and third harmonic generation (SHG and THG) or multiphoton excitation fluorescent emissions [8,9]. The use of MPM in mineralogy is, however, very limited. Multiphoton excitation of fluorescent emissions from gems and minerals has not been reported to date, to the best of our knowledge [2, *pp. 37*]. The first reported application of MPM in imaging of gems and minerals [10] showed only second and third harmonic generation signals. We have since been able to excite fluorescent emissions (two-photon excitation (2PEF) and three-photon excitation fluorescence (3PEF)) from a wide range of gems and minerals by using appropriate laser excitation wavelengths and filters in the UV and visible range. We observe both two-photon and three-photon excitation fluorescence in rocks, gems, and minerals from a wide variety of geologic settings. As a result, we can produce high resolution four-color images (SHG, THG, 2PEF and 3PEF) in 3D. This not only reveals the stunning beauty of the internal structures of minerals but can also provide unparalleled information regarding their chemical composition, origin, and formational histories.

## 2. Methods

The schematic diagram of our multiphoton microscope is shown in Fig. 1. A detailed description of the microscope has been published in [11]. The femtosecond laser that we used in this work was based on a Yb-doped fiber laser. The laser provides ~50 fs pulses, 8 nJ at full power and ~ 8 MHz repetition rate. The center wavelength of the laser is ~ 1040 nm. Different sets of optical spectral filters are used to acquire signal in various spectral windows. In general, we divide the luminescent signal emitted from minerals into 4 regions: **1.** SHG: 520 nm bandpass filter with 17 nm FWHM bandwidth; **2.** Two-photon excitation fluorescence (2PEF, this is just a nominal name since 3PEF can be in this region as well): 550-750 nm. **3.** THG: (340 nm bandpass filter with 26 nm FWHM bandwidth); **4.** 3PEF: 400-500 nm. We have only two photomultiplier tubes (PMT), so in order to acquire the 4 spectral regions we image the same area of the sample twice with two appropriate sets of filters. Signal from the two PMTs are acquired simultaneously. We have also been able to add a white light microscope port for ease of navigating across the sample surface and identifying locations of interest. This port is also used for spectral measurement with a spectrometer.

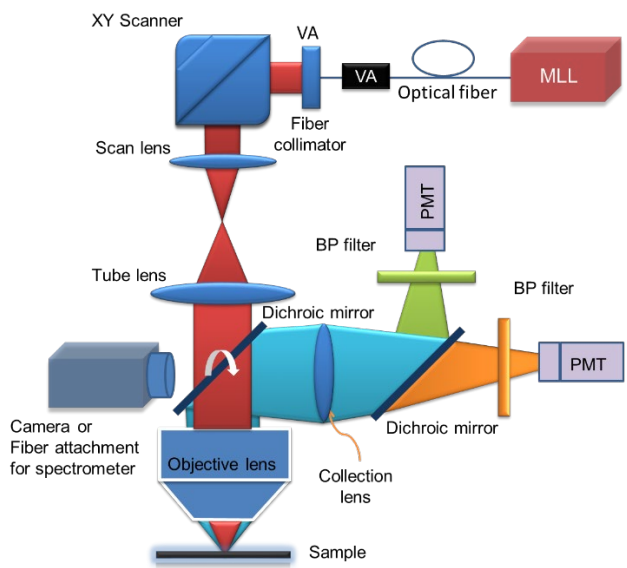

Fig. 1. Schematic diagram of our multiphoton microscope. A camera and fiber attachment port has been added to the microscope reported in [10,11] to record white-light microscopic images of samples and register the emission spectrum of the signals.

We used an Ocean optics QE6500 spectrometer to measure the luminescent spectrum in order to identify specific emission peaks measured <10 μm below the surface. These signals were then correlated with mineral phases identified at the sample surface with a Hitachi TM4000Plus tabletop scanning electron microscope (SEM) at the Kuiper-Arizona Laboratory for Astromaterials Analysis (K-ALFAA) at the University of Arizona. The SEM is coupled to an energy-dispersive x-ray spectrometer (EDS), which provides semi-quantitative chemical composition used for mineral identification.

Ideally, we would like to measure the spectrum for each pixel in the MPM image, but our assembly is not up to this task yet. Currently, we can only measure the emission from a small area of the sample since the scanning of the laser beam and the acquisition of the spectrometer is not synched. For that reason, we scanned the laser beam over imaged areas while the spectrometer was acquiring signal. Thus, the spectral information is acquired over a small, scanned area, and the resulting spectrum represents signals generated over the entire field of view (FOV).

**3.1 Fluorescent emission via nonlinear optical stimulation**

Most geologic samples show strong fluorescent emission along with the SHG and THG signal. Figure 2 shows a representative image from a roughly polished slab of dacite, a quartz-rich igneous rock, that we imaged with our microscope. Fig. 2.A, 2.B, 2.C, and 2.D correspond to the SHG, 2PEF, THG, and 3PEF, respectively. Figure 2.E is the combined image of all 4 channels (false color: red corresponds to SHG, green is THG, blue is the 2PEF channel, and cyan is the 3PEF channel). The SHG signal comes from crystalline structures that lack inversion symmetry, primarily the mineral quartz, as confirmed by compositional analysis with SEM. The THG signal is generated from interfaces when there is a change in the refractive index. The fluorescent signals (2PEF and 3PEF) are from fluorescent centers within the minerals. These can most likely be attributed to trace quantities of rare-earth elements (REE), anion vacancies, or a range of other established fluorescent centers in a mineral's crystal structure [2].

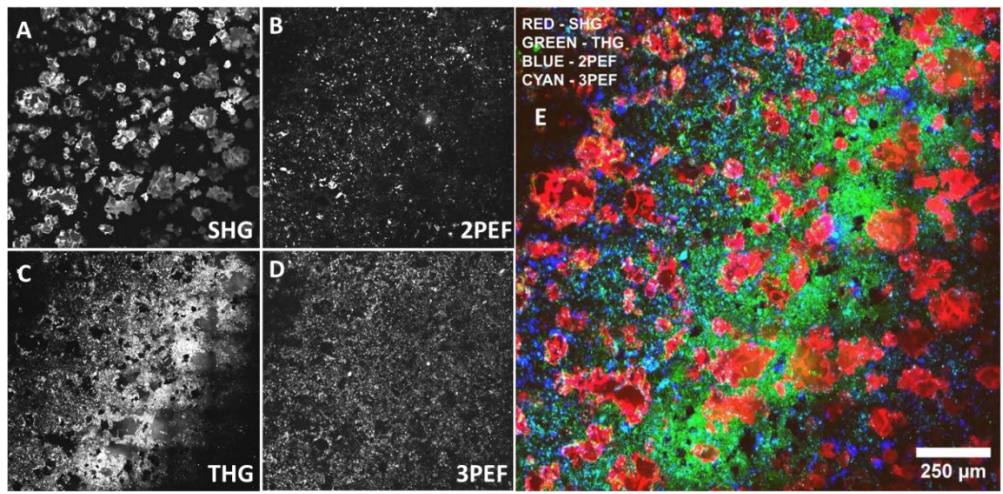

Fig. 2. Multiphoton image mosaic of dacite (~ 5 µm underneath the surface). A: SHG signal, B: 2PEF signal, C: THG signal, D: 3PEF signal, E: combined false colored image of 4 channels. SHG signals largely correspond to quartz, while THG signals are present in most other phases, including pyroxenes and feldspars. Fluorescence is most likely attributable to accessory fluorite (see text, Fig.3). The gradient of THG signal intensity in the mosaic is an imaging artefact due to uneven surface topography.

Figure 3 shows higher-resolution images and luminescence spectra collected from sub-regions of the dacite sample. Four distinct peaks were measured from region A. The THG signal is located around 350 nm. The 3PEF is near 440 nm; the center of this peak changed from location to location and from sample to sample. The SHG signal is located near 520 nm, as expected. The 2PEF signal is around 750 nm from regions A and B and 600 nm from region C. Given the correlation between 2PEF and 3PEF signals in various subregions, the 440 nm and 750 nm emission peaks are most likely attributable to the same component, the mineral fluorite ($CaF_2$). Fluorite is an accessory phase in this sample and is usually interstitial to major mineral phases: quartz, pyroxenes, and feldspars. Serendipitously, the indigo-blue emission of fluorite under UV-irradiation was also the first report of mineral luminescence by Stokes nearly 200 years ago, and is the origin of the term, "fluorescence" [1]. Likewise, the first experimental confirmation of two-photon excitation fluorescence utilized a synthetic fluorite crystal doped with $Eu^{2+}$ [12]. Both the 440 and 750 nm emission peaks in our measured spectra can also be attributed to the presence of trace quantities of rare-earth elements and anion vacancies in fluorite's crystal structure [2,12]. It is unclear what proportion of the 750 nm emission is produced by excitation at 347 nm (3PEF) or 520 nm (2PEF). Emission at ~750 nm can be stimulated by neighboring anion vacancies in fluorite (M-centers), in which two trapped electrons are excited by laser stimulation in the ultra-violet range [2]. In region C, a broad excitation peak is centered at ~620 nm. This peak likely corresponds to zircon ($ZrSiO_4$), another accessory phase identified with SEM. The activating source of this excitation feature in zircon is poorly understood but may be attributable to trapped electrons produced by the decay of radioactive $^{238}U$ [2]. This mechanism (i.e., trapped-state induced by radiation from the decay of radioactive elements) is widely observed in thermoluminescence studies [e.g., 3] and has also been observed by optically stimulated luminescence (OSL). OSL is a process in which a material, previously exposed to ionizing radiation, emits a light signal when subjected to appropriate optical stimulation [13]. The emitted optical signal is found to be proportional to the absorbed ionizing dose. The signals emitted via OSL are similar to those produced during thermoluminescence, where elevated temperature is used as the means for stimulation instead of a laser beam. The advantage of OSL is that it can induce luminescence in an irradiated sample many times whereas complete resetting is done in a single iteration of thermoluminescence. OSL has been successfully performed using CW or Q-switched lasers [13], but this is the first reported observation of radiation-induced luminescence using a femtosecond laser-pulsed source, which provides greater spatial resolution and better signal-to-noise than other methods.

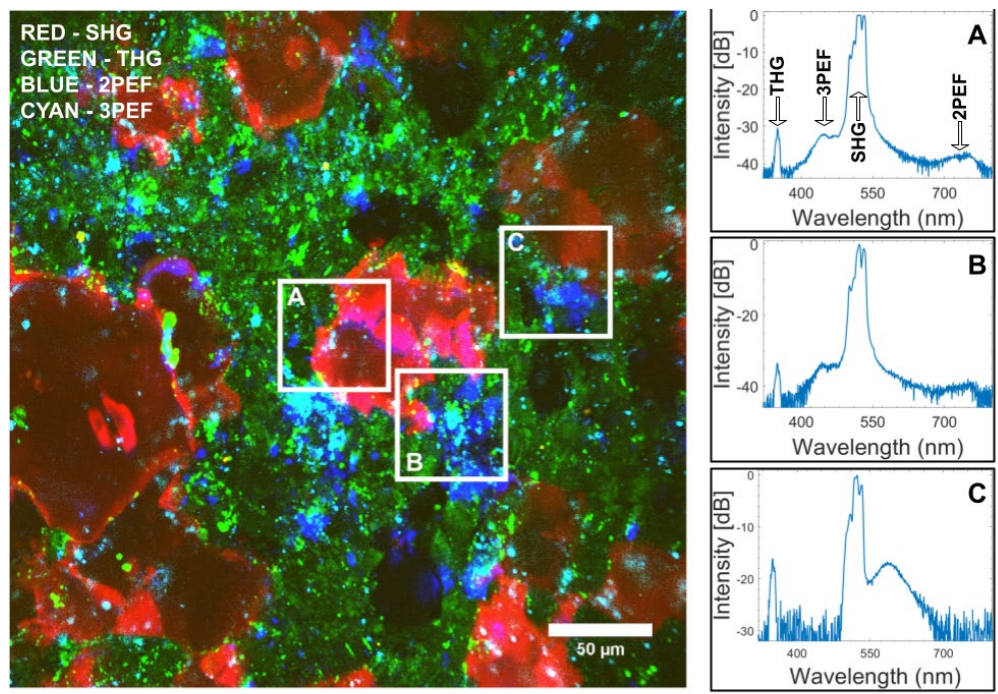

Fig. 3. Multiphoton image of dacite and emission spectrum from different areas within the sample. Areas **A** and **B** possess emission peaks at ~440 and 750 nm consistent with REEs in the mineral fluorite, $CaF_2$. Area **C** presents an emission peak at ~620 nm, possibly consistent with radiation-induced luminescence in the mineral zircon, $ZrSiO_4$.

We measured the signal power dependence on the excitation laser power to confirm the nature of multiphoton interaction as the mechanism of signal generation in our experiment. The result of the measurement is shown in Fig. 4. As expected, the power dependence of the SHG and 2PEF signals follows a quadratic function. The power dependence of the THG and 3PEF signals closely follows a cubic function.

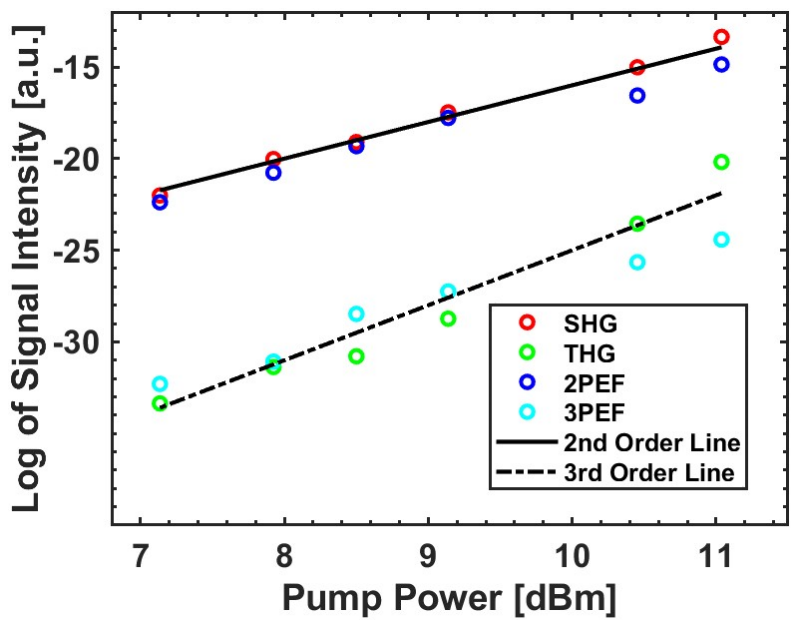

Fig. 4. Nonlinear signals generated from the Dacite sample as the function of the pump power.

Interestingly, we also observe photo-bleaching behavior of the fluorescent signals from minerals. This behavior is widely known for common fluorescent dyes or nanoparticles [e.g., 14]. Fig. 5 shows the nonlinear signals' intensity as the function of the number of scans. It is clear that the SHG and THG signals are stable under femtosecond laser irradiation. The 2PEF and 3PEF signals are, however, decreased over time as we scanned the focused laser beam over the same area multiple times. We observed that the rate of signal decay varies from location to location which may be a useful indicator to differentiate the different fluorescent centers within the mineral. We also note that the photo-bleaching behavior is also consistent with trapped states induced by ionizing radiation, which supports our interpretation that MPM can be used to excite radiation-induced luminescence.

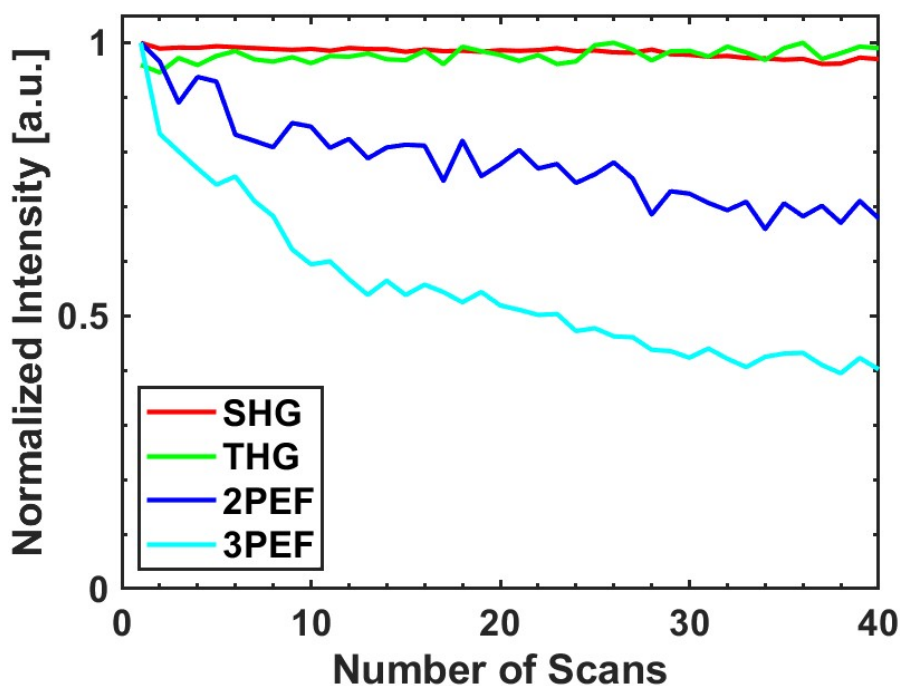

Fig. 5. Nonlinear signal intensities as the function of the number of scans.

We have been also able to perform 3D imaging of these samples. Figure 6 shows an example of such a measurement. We have imaged fluorescence to a depth of ~200 μm. The maximum depth of resolvable fluorescence depends on the optical transparencies of specific samples at relevant wavelengths and the nonlinear optical properties of overlying minerals.

### 3.2. Nonlinear optical imaging of fluid inclusions

As shown in Figs. 3 and 6, crystals often host smaller mineral grains included within them that are resolvable via MPM. These mineral inclusions can form in a variety of scenarios and provide critical information for interpreting the formational histories of gems, minerals, and host rocks. Primary inclusions become entrapped as a liquid or solid during the initial growth of the host crystal and preserve valuable information about the original composition of the medium from which a mineral crystalized. Secondary inclusions form after crystallization of the host mineral, providing records of subsequent alteration events. Inclusions can be solid or liquid, forming in both high temperature magmatic systems as well as low temperature aqueous fluid systems. As such, mineral and fluid inclusions are of great interest to geoscientists across a wide range of disciplines.

Usually, 2D petrographic analysis of fluid inclusions requires extensive sample preparation, including doubly polished thin sections, cryoanalysis, and other processing methods that damage or consume sample materials [e.g., 15]. We show that nonlinear optics can circumvent the need for such extensive preparation while providing equivalent and/or novel textural information. Of course, the MPM can also be used on prepared samples, which will provide even greater image resolution.

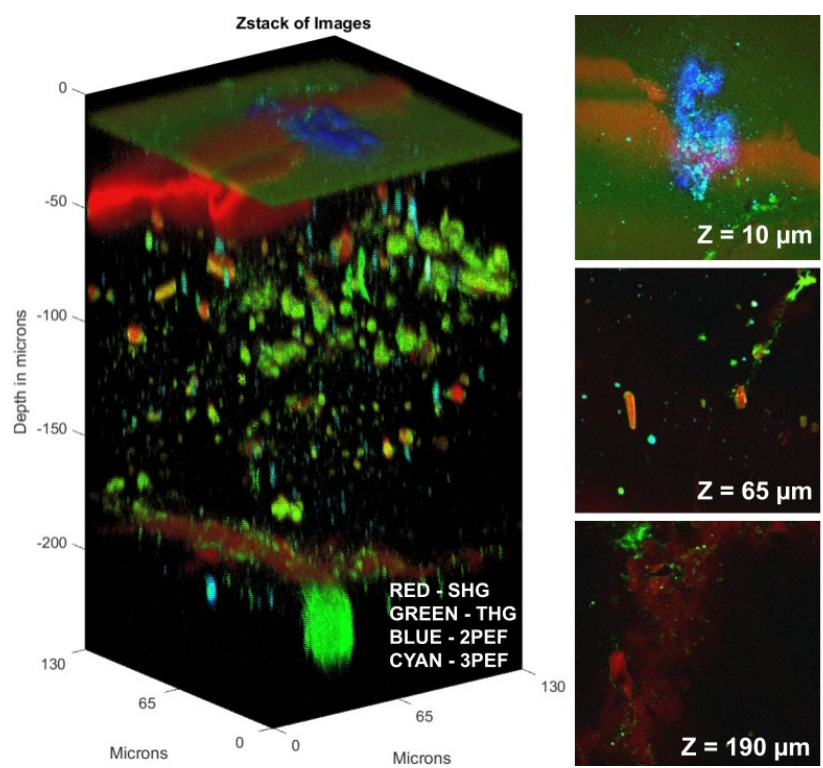

Fig. 6. 3D multiphoton profile for a sample of nepheline-syenite, a quartz-poor igneous rock. The blue fluorescence at the surface is likely a grain of sodalite. Numerous THG signals (~10 µm diameters) throughout the depth profile of the sample are likely small mineral grains included within larger host minerals. These images and 3D profiles can provide valuable petrographic context for the formational history of geologic samples, greatly expanding the information available from more traditional analytical methods that are often limited to a few micrometers below the sample surface.

Fig. 7 shows a 3D rendering of THG signals produced by a fluid inclusion in a single calcite ($CaCO_3$) crystal 50-70 µm below the sample surface. Live-capture video of THG signals from the interior of the fluid inclusion show pseudo-Brownian movement of a ~1 µm particle within the inclusion, most likely a bubble and/or precipitated salt crystal interacting with the laser (Online Supplements, Visualizations 1 and 2). This observation confirms that the inclusion contains fluid rather than an empty cavity or a solid mineral phase. A faint THG signal concentrically rims the exterior of the fluid inclusion, displaced by a few microns. This feature may represent a growth ring formed as the surrounding calcite crystal grew and entrapped the fluid. The upper and lower bases of the inclusion are flat with tapering edges. The vertical walls of the fluid inclusion are not visible due to the geometry of the optical beam waist. Otherwise, no THG or SHG signals were detected in the immediate vicinity (within 50 µm) of the inclusion, which could indicate that this is a primary fluid inclusion formed during initial crystallization of calcite [15]. No fluorescence was detected in this inclusion, although UV confocal scanning laser microscopy has been used to generate fluorescence in aqueous fluid inclusions containing organic molecules [15]. The same should be true for multiphoton fluorescence in fluid inclusions that bear organic molecules. It is likely that the long-UV wavelength produced by 3-photon interactions in our current assembly does not meet the minimum energy required to excite electrons from their ground state in organic compounds if any are present, but this can be addressed in future work through use of a laser source that can reach shorter UV wavelengths.

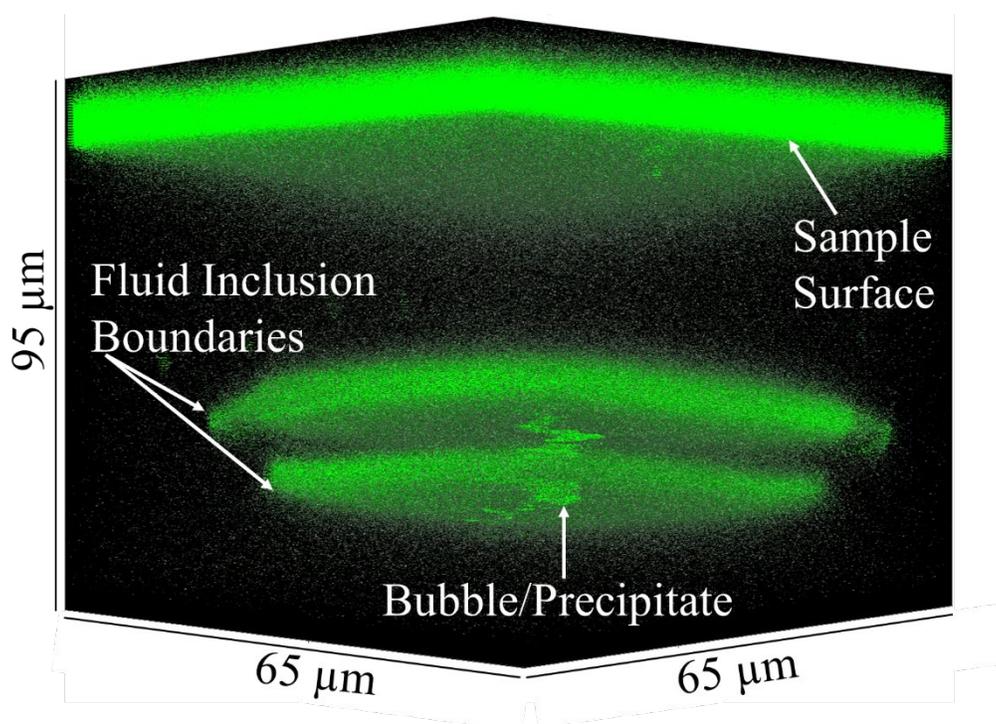

Fig. 7. THG signals (green) capture the 3D geometry of a ~45 µm-wide fluid inclusion 50-70 µm below the unpolished sample surface of calcite. Mobile bubbles and/or precipitated crystals were imaged during the 3D scan, producing a strong THG signal through the interior of the fluid inclusion along the vertical axis. Animated 3D models and live-capture videos are available in online Supplementary Materials (see Visualizations 1 and 2).

### 3.3 Applications of nonlinear optics to astromaterials

Nonlinear optical mineralogy using multiphoton microscopy is currently an untapped resource in geosciences and related fields, but our initial findings suggest that it has the potential to revolutionize microanalysis of geological materials across a wide range of topics, as it has already done with tissue analysis in fields of biomedical research. Due to its nondestructive nature, immediate applications may include investigation of particularly rare or unique samples, including meteorites and samples returned from other planetary bodies via spacecraft (i.e., astromaterials).

As a first-order assessment of this tool's capabilities for investigating astromaterials, we imaged a polished section of the meteorite Northwest Africa (NWA) 10421, a rare variety of primitive asteroidal meteorite classified as a thermally altered Rumuruti-type (R6) chondrite. Such samples are mineralogically complex, and a detailed description of its mineralogy and dynamical origin is beyond the scope of this work. Rather, we report initial observations of nonlinear optical signals and suggest possible applications to future targeted investigations of astromaterials.

Second and third harmonic generation signals are ubiquitous throughout the sample. Faint SHG signals manifest near most grain boundaries, while THG is produced at grain interfaces with contrasting refractive indices. The strongest SHG signals are produced by chlorapatite, $Ca_5(PO_4)_3Cl$. Its strong SHG signal is due to a lack of inversion symmetry in its monoclinic crystal structure, which indicates that the sample slowly cooled to temperatures below 200 °C [16]. In the largest chlorapatite (Fig. 8), which preserves a roughly hexagonal geometry, THG manifests perpendicular to the longest axis, possibly corresponding to a weak basal cleavage plane. Additional THG signals are also superimposed onto chlorapatite with smaller, roughly hexagonal geometries (~40 μm diameters), possibly indicating that chlorapatite began to morph to its higher symmetry form during impact-induced thermal alteration.

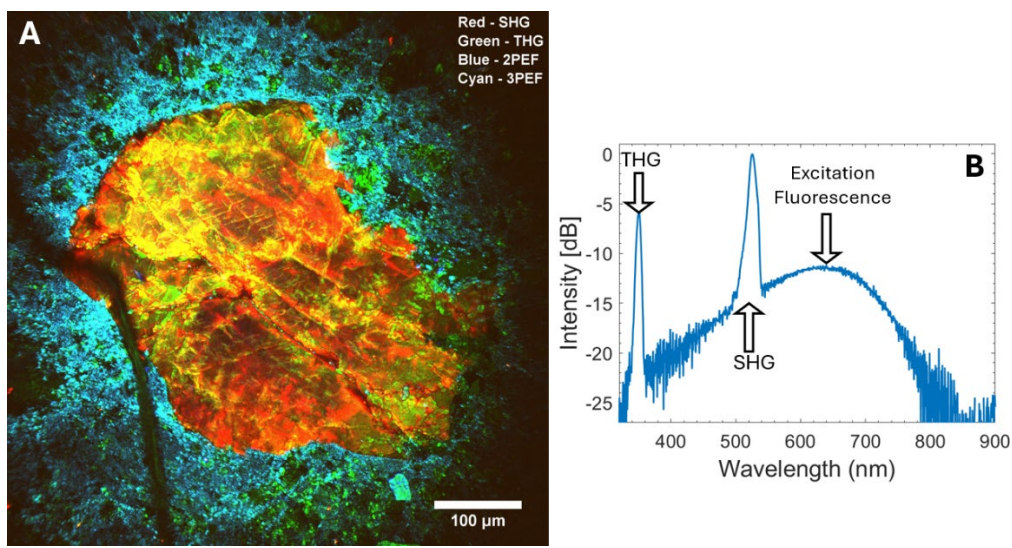

Fig. 8. (A) Four-channel image showing harmonic generations and multiphoton excitation fluorescent channels from the R6 meteorite Northwest Africa (NWA) 10421. The mineral chlorapatite (center of image) produces strong SHG signals (red) but fluorescence is dominated by surrounding minerals, primarily olivine. The luminescence spectrum (B) for this region shows a broad excitation feature centered at ~650 μm, consistent with radiation-induced luminescence of olivine [20].

Intriguingly, NWA 10421 exhibited strong 2PEF and 3PEF signals across most of the sample, particularly from the mineral olivine, $(Fe,Mg)_2SiO_4$, which is the most abundant phase in R-type chondrites (>60 volume %, [17]). Fluorescence of olivine was 2 orders of magnitude greater than measured luminescence in terrestrial dacite. Such strong fluorescence from R6 meteoritic olivine was unexpected, as its high $Fe^{2+}$ content (~10 mole % [e.g., 18]) should completely quench linear forms of luminescence [3]. Olivine luminescence observed with MPM may be attributable to the exponential relationship between the atomic transition rate of excitation and the intensity of the laser in nonlinear optical interactions [19]. Consequently, the rate of fluorescent emission overwhelms the quenching rate of $Fe^{2+}$, yielding luminescence that is not otherwise resolvable via methods that employ linear optics (e.g., thermoluminescence, OSL). Olivine produces a broad emission peak centered at ~650 nm (Fig. 8b). This peak coincides with thermoluminescence emissions of experimentally irradiated Fe-poor olivine [20]. The mineral pyroxene, $Ca(Mg,Fe)SiO_3$, shows no resolvable fluorescence. This observation is also consistent with experimental irradiation of pyroxene [20]. Other mineral phases appear to show some degree of fluorescence in images, but their individual excitation peaks are overwhelmed by signals generated from olivine and difficult to distinguish in the FOV-averaged spectra. The phases likely contribute to the tail of emission at higher wavelengths. These phases include plagioclase feldspar $((Na,Ca)(Si,Al)_4O_8)$ and pyrrhotite $(Fe_{1-x}S)$. Feldspars have been reported as the main source of radiation-induced thermoluminescence in meteorites [e.g., 21], although not all feldspars in our sample fluoresce. This may be due to partial melting of some feldspars during impact, which would have released the trapped electrons responsible for radiation-induced fluorescence. Notably, pyrrhotite only produces fluorescent signals in regions of the sample showing mottled, porous textures that are similar to pyrrhotite in other aqueously altered meteorites (Fig. 9, compare to Fig. 5 in [22]), suggesting that pyrrhotite fluorescence may be attributable to a water-rock interactions. The combined ability to image fluorescence of aqueously altered sulfides and fluid inclusions in carbonates could be of great interest to investigations of samples returned via the OSIRIS-REx spacecraft from the asteroid Bennu, which preserves evidence of aqueous fluids from the early solar system [23].

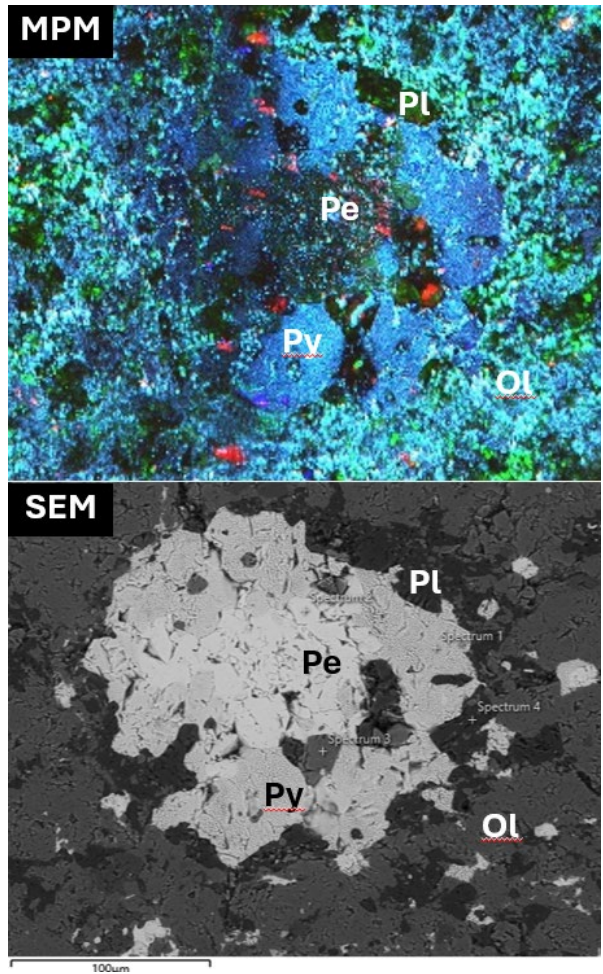

Fig. 9. Four-channel multiphoton image (top) shows fluorescence of iron-sulfide (pyrrhotite), but only for areas with mottled, porous textures in SEM images (bottom) that may be correlated with aqueous alteration via a fluid medium. Abbreviations label specific minerals in each image: Py = pyrrhotite, Pe = pentlandite, Pl = plagioclase feldspar, and Ol = olivine.

At least two potential sources may contribute to radiation-induced luminescence of olivine in meteorites, and both are testable by well-established methods in astromaterials analysis. The first is cosmic ray exposure (CRE), which occurs after a meteoroid fragment has been excavated from the interior of its parent body by collisional impact. Exposure to cosmic radiation alters the electronic structure of minerals, leading to fluorescence when stimulated, and it also imparts isotopic anomalies. The magnitude of these anomalies is proportional to the amount of time that the meteoroid was exposed to cosmic irradiation while in transit through the solar system. Thus, a coordinated study of radiation-induced luminescence with measured CRE ages would lend further insight into the relationship between the cosmic irradiation and luminescence. Alternatively, radiation-induced luminescence can also be caused by the decay of unstable radionuclide isotopes [e.g., 2]. In the early solar system, the most abundant short-lived radionuclide was $^{26}$Al. The energy released during the decay of $^{26}$Al is generally accepted as the heat source that drove thermal alteration and internal igneous differentiation (i.e., melting and structural stratification) of planetary bodies across the solar system, although a demonstrative causal relationship between the two has proved elusive and some models continue to argue for other heat sources as major contributors to internal planetary differentiation [summary provided in 24]. A coordinated investigation into radiation-induced luminescence, mineralogic thermal alteration, and the isotopic signatures of short-lived unstable isotopes could provide the most direct evidence for the role of radionuclide decay in planetary differentiation.

Regardless of the source of radiation, this sample also shows clear evidence for break-up and reassembly of disparate mineralogic components from different sources, which is also manifested in the contrasting luminescence between the host mineralogy and exogeneous fragments (Fig. 10). It is clear that the host mineralogy was irradiated more extensively than the incorporated clastic fragments. Given the high degree of thermal alteration in the strongly luminescent host mineralogy, our preferred interpretation is that luminescence in this sample is primarily related to radionuclide decay. However, such interpretations must be made in the petrographic context of specific samples. Spatially resolved measurements of $^{26}$Mg (the decay product of $^{26}$Al) could confirm the source of irradiation in NWA 10421 and other samples. Similarly, measurements of isotopes imparted by cosmic rays could distinguish between the two potential sources of radiation. We defer such methods to future targeted investigations.

4. **Conclusions**

We have demonstrated that nonlinear optical mineralogy, as facilitated by multiphoton microscopy, holds significant potential for new avenues of scientific inquiry across a broad range of geoscience and planetary materials applications. For the first time, we report multiphoton-stimulated excitation fluorescence in minerals, which provides a new method of measuring luminescent centers in minerals and imaging microscale textural features. We also report the first integrated use of harmonic generations and nonlinear induced fluorescence to investigate 3D microstructures of mineral and fluid inclusions within rock samples. Finally, we report the first use of nonlinear optics on astromaterials and have found that this tool can provide unique opportunities to measure radiation-induced luminescence that may have been imparted by cosmic ray exposure and/or the decay of radionuclide isotopes. Further development of methods in nonlinear optical mineralogy is ongoing and shows potential to revolutionize the analysis of geologic samples and astromaterials akin to the impact that multiphoton microscopy has had on biomedical research over the last two decades.

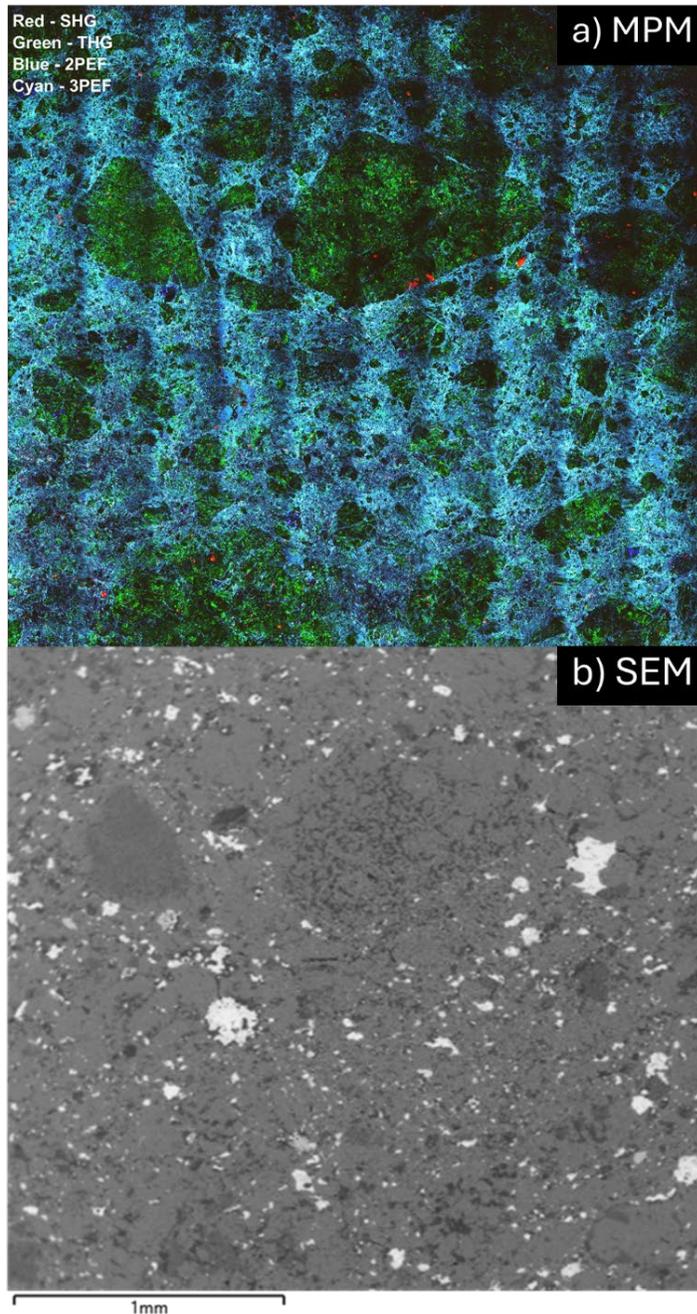

Figure 10. MPM mosaic (a) across a 2.5 mm subsection of the meteorite NWA 10421. Multiphoton fluorescence (blue/cyan) is ubiquitous throughout the most matrix of the specimen but is absent from fragmental clasts (green) that were most likely incorporated during reassembly after catastrophic impact. Evidently, the olivine, feldspars, and sulfides in these non-luminescent regions did not experience the same radiation history as the host matrix, providing critical insight into the formational history of the meteorite. When contrasted with an SEM backscattered electron image (b) from the same region, it is clear that the nonlinear optical imaging of meteorites offers novel information for astromaterials that is not readily accessible using conventional imaging methods.

**Acknowledgments.** The authors wish to thank P. Haenecour for providing tabletop SEM access to SDC for mineral identification.

**Funding.** Funding for SDC was provided in part by the Technology Research Initiative Fund (TRIF) program at the University of Arizona.

**Data availability.** Supporting data for figures in this work can be accessed via the The University of Arizona Research Data Repository (ReDATA). DOI: 10.25422/azu.data.26506453. Prior to acceptance, reviewers may access the archived data at https://figshare.com/s/443798f943ce3fea47dd.

**Supplemental document.** Supplementary visualiztions are available in the online appendix.